# The Evolution of the DELTA Control System

E. Kasel, B. Keil, D. Schirmer, D. Zimoch
Institute for Accelerator Physics and Synchrotron Radiation,
University of Dortmund, Germany

*Abstract*

Since January 1999 the DELTA control system is changing from a non standard hand-made system to the standard Experimental Physics and Industrial Control System (EPICS) [1]. The complete 80 MeV LINAC and transfer-line, the asymmetric 5.5 T multipole wiggler magnet (SAW) and main parts of the 1.5 GeV electron storage ring are now under EPICS control.

Additional hardware upgrades like a fast 100baseTX/FX Ethernet network, Linux-PCs, PowerPC-IOCs, and reengineered device I/O-interface cards have been installed. Moreover, a self-made VME DSP-board is in the final test phase to control the booster ramping process. Generic stream device drivers for the IEEE/GPIB and CAN fieldbusses have been developed and are already in use since one year [2].

The number of high level software tools for data acquisition and evaluation as well as for remote device control and complex software feedbacks is increasing continuously. The availability of a wide variety of interfaces to the EPICS Channel Access (CA) protocol makes it possible to include state of the art software platforms and architectures like the CORBA middleware, Java, Delphi, Tcl/Tk etc. very easily. Furthermore, the integration of complex simulation, design and modeling tools together with database archiving and Web-based accelerator monitoring is in progress.

## 1 INTRODUCTION

During the last years the self-made DELTA control system has been replaced by EPICS. The change had to be done step by step because storage ring operation had to be ensured during the migration. In summer 2001 this migration was finished. The whole facility, the 80 MeV linac, the 1.5 GeV booster, the storage ring and the Superconducting Asymmetric Wiggler (SAW), is now handled by EPICS.

Combined with the migration of the control system exchanges of VME computer types (from Force- and Microsys-CPUs to Power-PC) and upgrade of the internal Ethernet network to 100 MBit and fiber optics had been made.

## 2 REACHING EPICS

The DELTA control system has the usual three level soft- and hardware structure consisting of fieldbus systems which are linked via 28 VME-IOCs and local network to the front-end computers for high level operator interfaces. Altogether, there are 46 computers in the control network.

Because many devices had to be accessed by serial protocols we developed a generic driver which can be configured by an plain text file. Due to the modular structure (see fig. 1) of this driver new EPICS records and also new fieldbus systems can be applied easily. The flexibility of the driver leads to its use for 672 of the 4767 records (process variables in EPICS) in the DELTA control system (see table 1).

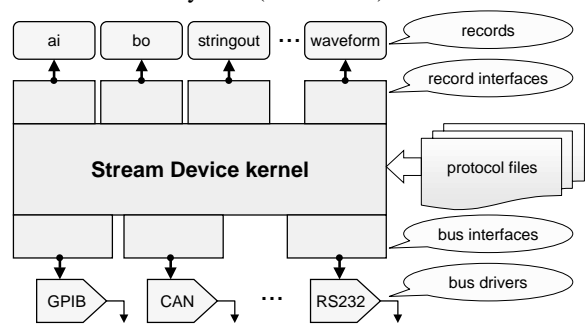

Figure 1: Structure of Stream Device Driver

HP-Workstations as operator interfaces are replaced by Linux-PCs. At the moment, there are still several workstations in use, but mainly as display computers for programs running on the PCs.

| Device Type | Number of Record |
|---|---|
| Soft Channel | 1909 |
| Stream | 672 |
| Delta DSP | 561 |
| CANbus | 497 |
| Delta-FG | 490 |
| Desy CPS | 244 |
| PPIO Chopper | 222 |
| Others | 172 |
| **Total** | **4767** |

Table 1: Records in the Delta Control System

For the reengineering of the display panels the EMW-GUI-builder was developed so that even complex graphical user interfaces (GUI) can be generated without programming knowledge. The whole front-end programming was done on the basis of Tcl/Tk. Even extensive applications like beam based calibration (BBC) routines and online optic measurements are written with this toolkit.

## 3 ORBIT DRIFT FEEDBACK

On a slow time scale, DELTA shows significant orbit drifts. Two main reasons could be detected: thermal effects and field drifts of the Superconducting Asymmetric Wiggler (SAW). Thermal effects are beam current related and caused by the heat load from synchrotron radiation. The changing temperature leads to deflections of the vacuum chamber. This causes displacements of the quadrupole magnets [3].

While the Superconducting Asymmetric Wiggler (SAW) is operative, its slowly increasing field error is the dominating reason for orbit drifts. This field error is caused by different current losses in the three separate circuits of the SAW [4].

Both effects could be successfully suppressed by a EPICS-based feedback system implemented in Tcl/Tk on a Linux PC. Orbit correction is performed once a second in both planes. The machine operator can choose from a variety of different correction algorithms. All correction methods work on measured beam response matrices, which have proved to be a better basis than calculated models [5].

Currently implemented methods are:
- most effective corrector
  This chooses the steerer that will minimize the rms orbit error.
- most efficient corrector
  This chooses the steerer that needs the smallest change of steering to correct the rms orbit error.
- weighted correction
  This chooses a steerer that will correct the rms orbit error while reducing the absolute steerer strength.
- local bumps
  Three consecutive steerers are used to create a local orbit bump that reduces the largest orbit displacement.
- SVD (singular value decomposition)
  SVD is a numerical algorithm to 'invert' non-square or singular matrices. Applied to the beam response matrix, it allows to use all steerers to reduce all orbit displacements simultaneously.

The feedback system uses 43 beam position monitors, 30 horizontal and 26 vertical steerer magnets. The range of the steerers is ±3.1 mrad in the horizontal plane and ±1.1 mrad in the vertical plane. The resolution is $2.4 \cdot 10^{-4}$ of the maximum value. Due to the serial communication interface of the steerers, corrections can only be done slowly.

A correction rate of once a second has emerged to be entirely sufficient to stabilize the orbit on the required time scale. Removal of high frequent orbit oscillations is beyond the intention of this application and needs a separate set of fast steerers and faster hardware for acquisition and processing of orbit data.

## 4 THE BOOSTER DSP NETWORK

The 1.5 GeV booster BoDo is actually a ramped storage ring, with an energy range of 35 to 1500MeV and typical ramp cycle times between 2 and 7 seconds, depending on injection and extraction energy.

In May 2001, VME low level software and GUIs were changed from a self-made control system to EPICS, still using the old hardware: simple synchronized digital function generators VME boards (FGs) with external DACs for ramped power supply control, digital I/O VME boards, and CANBus modules with digital or 12 bit analog I/O that provided time stamp precision of about 100ms and readout frequencies of some Hz.

Since October 2001, all ramped booster magnet power supplies (dipole, 6 quadrupoles, 2 sextupoles, steerers), rf power, and all booster diagnostics (beam position measurement, tune control, beam loss measurement, power supply currents, ...) are handled by VME DSP boards developed at DELTA ("DeltaDSP"). New self-made standalone eurocards with DC-isolated temperature-stable 16 bit ADCs and DACs and fast interfaces to the DeltaDSP boards were installed, with minimum distance to power supplies resp. signal sources. Up to 8 DAC cards and 128 ADC cards can be connected to each DSP board, with data transfer rates up to 32 MByte/s for DAC cards (16 bit parallel bus, 10 m max. cable length) and 10 MBit/s for ADC cards (serial daisy chain, 50 m max. cable length). All booster diagnostics data is now available at ADC sample rates between 10 kHz (BPMs) and 100 kHz, with 300 ns time stamp precision.

Each DeltaDSP board has two 32 bit floating point DSPs (Analog Devices ADSP-21062), 3 MByte fast SRAM program memory, up to 64 MByte DRAM data memory and 1 MByte VME dual ported memory. External devices can be connected to four synchronous serial interfaces (32 MBaud max.) or two CANBus master interfaces (1 MBaud max.). Furthermore, a general-purpose flatwire differential I/O port is provided. Reconfiguration of an FPGA connected to this port provides flexible high-speed interfacing to different accelerator devices like DACs, DDS function generators or switched power supplies with inputs for digital current control via pulse width modulation.

Up to 255 DeltaDSP boards can be connected to a multiprocessor cluster using "DeltaNet", a novel FPGA-based 160 MBaud real-time field bus developed at DELTA. DeltaNet provides real-time data transfer with predictable bandwidth and bus access time, software-independent trigger distribution and accelerator-wide clock synchronization that tolerates transmission errors. Each DeltaNet node is connected to its next neighbour node by one fiber optic cable (max. length 2km node to node, 200km overall). The synchronous hardware design allows future DeltaBus implementations with much higher baud rates. ICs for a 1.25 GBaud implementation are already available.

The DeltaNet bus protocol and interface are completely implemented in hardware (FPGA), including 32 bit CRC checksums and state machines for power-up, bus arbitration and error handling. This solution provides maximum speed (1.5 μs package routing time, independent of data size) at minimum software overhead. Package transfer between DSP and DeltaNet FPGA uses DMA. All DeltaNet nodes receive all data packages transmitted by all other nodes.

The DSP boards have a generic EPICS driver/device support, i.e. the same driver/device support is used for DSP boards with different firmware and FPGA configurations.

Before installing the DSP system, all parameters of the booster ramping process had to be optimized manually, which usually took many hours. Since the control system could not provide most diagnostics data with sufficient time resolution and precision, reasons for bad booster performance were hard to detect.

Since the DSP system makes all diagnostics data available to all DSP boards via DeltaNet with high speed and precision, it is now possible to automatize machine operation by implementing intercommunicating distributed real-time control loops up to some 10 kHz bandwidth, e.g. fast orbit feedback, digital power supply current control and real-time tune and beam optics correction [6]. Furthermore, the DSP system can be adopted for a fast global orbit feedback in the main storage ring.

## 5 DEVELOPEMENT OF AN AGENT SYSTEM

An agent system consists of several independent programs which can communicate with each other and which perform their task highly autonomous. The main advantages of an agent system are the high flexibility and the possibility to add new agents to meet additional requirements of the system [7]. Both features are needed in the DELTA accelerator facility: on one hand easy handling and if possible automatic control for routine operation is required, on the other hand we need a flexible toolbox for accelerator experiments.

The knowledge base of the planned agent system is splitted into two parts: the actual machine values are handled by EPICS and for the more static data, like magnet positions, an Oracle database was installed. Most of the agents will access at least one of these sources. In addition, each agent has the ability to ask the other agents for information via TCP socket connection.

For example, the alarm handler can provide information about actual errors. The display-agent can ask for this errors and determine the physical position of the underlying hardware with the help of the static database. This combined information can be displayed on a map representing the machine and so the responsible technician can find the faulty devices (avoiding confusion about different naming conventions in the technical staff and the control system).

At the moment the design requires six agents: a manager-agent for debugging and control of the running agent system, an alarm handler which can detect faults and display errors in natural language, a status-agent for valuation and display of the actual machine state, a display-agent to provide information about hardware positions, help texts, naming conversions and technical terms, a program handler which supports the locating and execution of programs of the control system, and last but not least an expert system for semi-automatic control of routine operation.

## REFERENCES

[1] D. Schirmer, et al., Standardization of the DELTA Control System, Proceedings of the 7th ICALEPCS 1999, Trieste, Italy
[2] www.delta.uni-dortmund.de/controls/pub/doc
[3] G. Schmidt et al., Position Monitoring of Accelerator Components as Magnets and Beam Position Monitors, DIPAC 2001, Grenoble
[4] D. Schirmer, et al., First Experience with the Superconducting Asymmetric Wiggler of the DELTA Facility, EPAC 2000, Vienna
[5] D. Zimoch, Ph.D. thesis, to be published
[6] B. Keil, Ph.D. thesis, to be published
[7] G. Weiss, Multiagent Systems, MIT Press 1999